\documentclass[11pt]{article}
\usepackage {amsfonts}
\usepackage {graphicx}
\usepackage {epsfig}
\usepackage{amssymb}
\usepackage{amsmath}
\usepackage {longtable}
\usepackage[english]{babel}
\begin{document}

\title{TECHNOLOGY FOR PRODUCTION OF ULTRA-THIN
CRYSTAL SILICON MEMBRANES}

\author{V.B. Vyssotsky$^1$, E.V. Lobko$^2$, A.S. Lobko$^3$ \\ \\
{\it $^1$ ''Integral'' Production Concern, Minsk, Belarus}\\
{\it $^2$ Research Institute for Radio-materials, Minsk, Belarus} \\
{\it $^3$ Research Institute for Nuclear Problems, Belarus State
University, Minsk, Belarus} \\ {\it lobko@inp.minsk.by}}

\maketitle

\begin{center}
\begin{abstract}
Ultra-thin crystal targets in shape of self-supporting membranes
were produced for our experiments with parametric x-rays emitted
by low-energy electrons. Concise description of the technology
developed for production of silicon crystal membranes and the
technique for the membrane thickness measurements are considered
in the paper. Membrane thickness of $\sim $0.5$\mu$m at 1.0mm
diameter supporting by 2$\times$2 mm silicon substrate of $\sim
$200$\mu$m thickness is achieved. Developed technology for thin
crystal membrane production can also be applied for manufacturing
of various membrane-containing sensors.
\end{abstract}
\end{center}

Experimental research with parametric x-rays emitting by
non-relativistic electrons (NPXR) \cite{1} demands self-supporting
crystal targets (membranes) of high perfection with thickness of
0.5micron and below. Concise description of the technology for
production of ultra-thin crystal membranes developed for silicon
and the technique for the membrane thickness measurements are
considered in the current paper.

Typical targets for NPXR generation were made as of 2$\times$2 mm
silicon substrate of $\sim $200 $\mu$m thickness with crystal
membrane of 1.0mm diameter and $\sim $0.5µm thickness located
inside this support. Base planes have (100) or (111) orientations.
For clarity development of (100) membranes are discussing below.

Membrane material is the layer of un-doped epitaxial silicon of
$\sim $0.9-1.0$\mu$m thickness grown on substrate of boron doped
p-type Si with 0.01Ohm$\cdot $cm resistivity of KDB 0.01<100>
grade. Choice of such structure was determined by applied
electrochemical etching technique, in which un-doped epitaxial Si
serves as termination (or stop layer). To obtain membrane of other
thickness one should take structures with epitaxial layer
thickness close to the desired one. Precise membrane thickness
adjustment can be performed by ion-beam etching with rate $\sim
$10-15nm/min, but ion treatment for a long time leads to the
significant membrane surface erosion.

To perform electrochemical etching we produce electrical contact
to KDB $0.01<100>$ through high-resistive epitaxial Si layer from
the planar side and reduce substrate thickness to $\sim
$200$\mu$m. Substrate thinning is performed by mechanical
polishing. For contact manufacturing planar side is covered by
chemical resistant varnish with followed opening of the substrate
peripheral region. Then the substrate is etched in Si etchant to
the epitaxial layer ($\sim $1$\mu$m) depth (Fig. 1a).

After varnish stripping and appropriate chemical treatment,
non-planar side is covered by SiO$_{2} $of about 0.3-0.4$\mu$m
thickness, then Si$_{3}$N$_{4}$ of $\sim $0.15-0.2$\mu$m
thickness, and at last Cr layer of $\sim $0.05$\mu$m is vapor
deposited to the better photoresist adhesion \cite{2} (Fig.1b).
Next step is the membrane photolithography on the non-planar wafer
side. Electric contact to KDB $0.01<100>$ is provided by vapor
deposition at Ò$\sim $180$^{\circ} Ñ$ of $\sim $0.5$\mu$m aluminum
film through the mask to the substrate periphery, Fig.1c.

The membrane etching is performed in two stages. In the first
stage we carry out electrochemical porous anodic treatment in
HF:Ñ$_{2}$H$_{5}$OH=4:1 electrolyte \cite{3} during $\sim $120
minutes at $\sim $1.8$\mu$m/min rate. Process is going until it
reaches epitaxial Si at $\sim $200$\mu$m depth (Fig.1d). Finishing
of porous silicon formation one can observe at $\sim $10-20\%
potential jump. Additional etching during $\sim $10min is needed
for porous Si levelling along total substrate surface and also for
membrane thinning down to $\sim $0.5$\mu$m. Etching rate of
high-resistive epitaxial Si is low, approximately 50nm/min.

In the second stage porous Si is etched in 1\% KOH solution during
30minutes (Fig.1e). After porous Si etching the substrate is
extracted out of holder. Photoresist, Cr, and Al are stripped in
standard etchants. Dielectric (SiO$_{2}$+Si$_{3}$N$_{4}$) film is
stripped by plasmachemical treatment.

Chipping is carried out by disk cutter while substrate is
naphthalene glued on supporting Si wafer. Finished crystals
(Fig.1f) with membranes come out after naphthalene sublimation in
the thermostat at $\sim $110$^{\circ} Ñ$ temperature.

Final thickness of membranes is of the 0.4-0.9$\mu$m range. To get
thinner membranes one should apply ion-beam etching. Our
experience shows that membranes with thickness below $\sim
$0.2-0.3$\mu$m can be hardly obtained in self-supporting mode due
to its un-sufficient mechanical strength.

We have developed relatively simple technique to measure thickness
of such ultra-thin Si targets. As Si membranes with thickness of
about micron and thinner are semi-transparent in the visible
light, one can record their optical transmittance spectra (example
is in Fig.2, right panel). We can see here interference maxima
against background of a standard optical transmittance. After
background subtraction precise values of maxima locations can be
measured (Fig.2, left panel).

Taking into consideration Si refraction index dispersion (it has
about 10\% variance in the visible region) and combining
interference conditions for neighboring maxima one can obtain
following expression for thickness of a parallel-sided plate

\[
d = \frac{{\lambda _{i} \lambda _{i + 1}} }{{2\left( {n_{i + 1}
\lambda _{i} - n_{i} \lambda _{i + 1}}  \right)}},
\]

\noindent where $n$ is the refraction index, $\lambda$ is the
wavelength in a maximum, $i$ and $(i+1)$ identify values belonging
to adjoined interference maxima. Data on some measured samples
took randomly of a production lot are presented in the Table.
Dispersion of measured thickness values of five targets took
randomly from one lot is below 5\%.

Developed technology for thin crystal membrane production can also
be applied for manufacturing of various membrane-containing
sensors, e.g. pressure meters, gas concentration and flow sensors,
as well as other micro-mechanics items.

\begin{figure}
\epsfxsize = 15 cm \centerline{\epsfbox{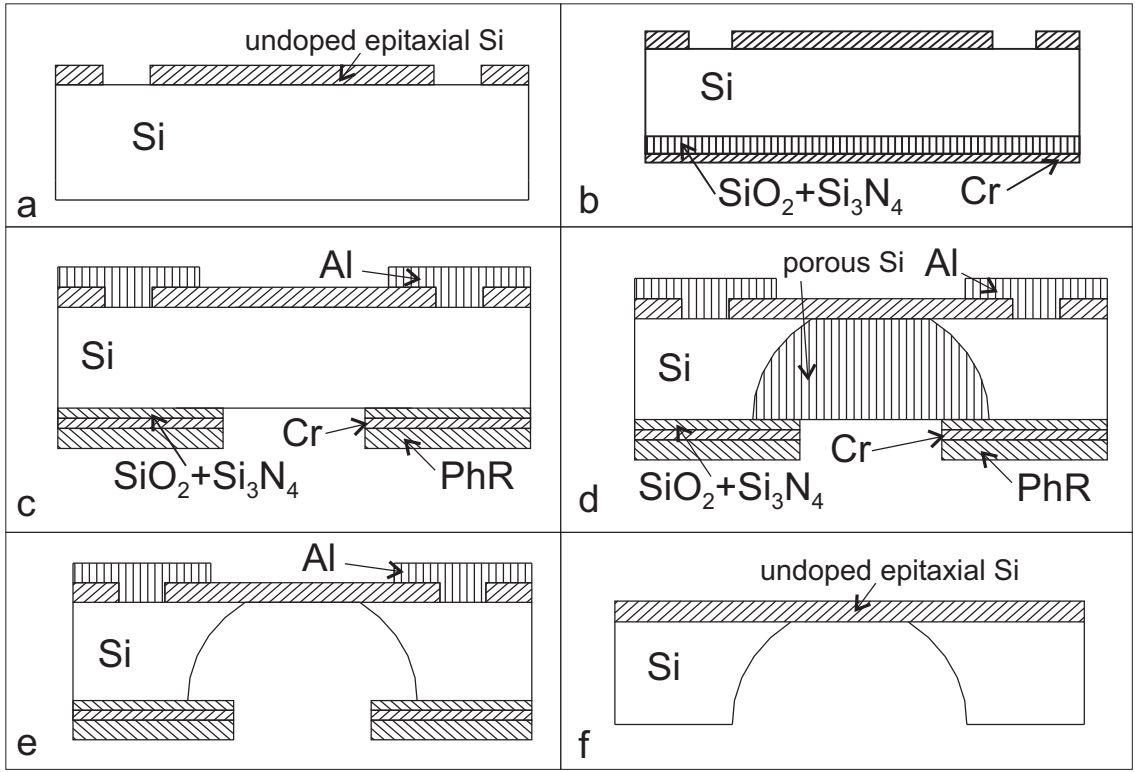}}
\caption{Technology route for crystal Si membrane manufacturing}
\end{figure}

\begin{figure}
\epsfxsize = 15 cm \centerline{\epsfbox{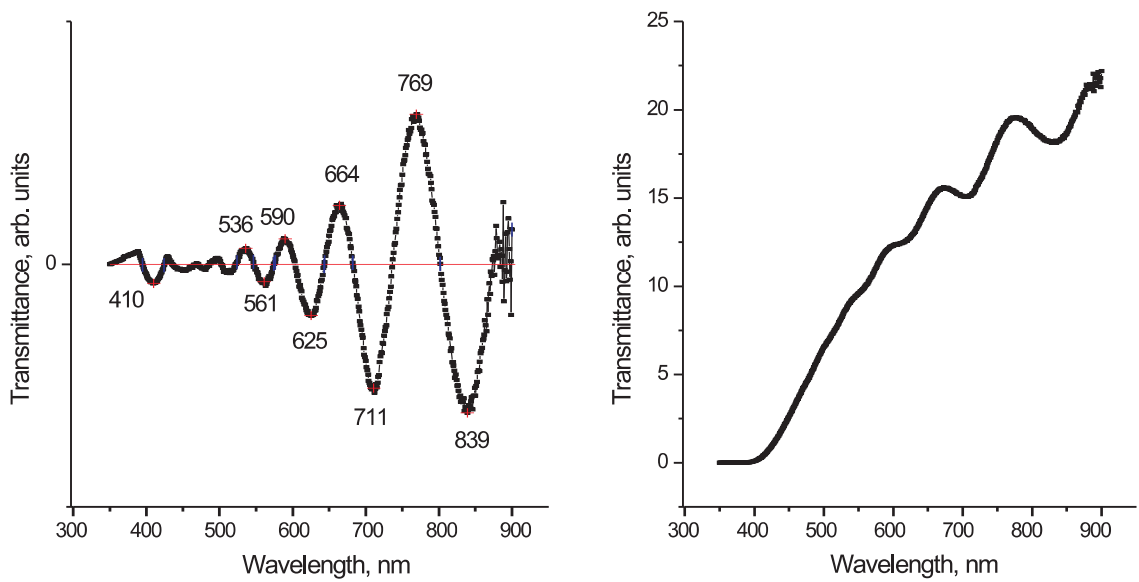}} \caption{Target
optical transmittance (right) and interference maxima locations
(left)}
\end{figure}

\begin{table} \centering \caption{Results of target thickness
measurements}
\begin{tabular}{cccccc} \hline\noalign{\smallskip}
\hline Sample No& 1& 2& 3& 4&
5 \\
\hline
\\
Thickness, nm& 520$\pm $2& 514$\pm $3& 505$\pm $25& 495$\pm $17&
467$\pm $25 \\
\hline
\end{tabular}
\end{table}



\begin{thebibliography}{999999}

\bibitem{1} V.G. Baryshevsky et al. LANL e-print arXive physics/0507036

\bibitem{2} A. Splinter, O. Bartels, W. Benecke // Sensors and Actuators
B76 (2001) 354-360.

\bibitem{3} G. Lammel, Ph. Renaud // Sensors and Actuators 85 (2000)
356-360.


\end{thebibliography}
\end{document}